\documentclass{PoS}

\usepackage[varg]{txfonts}
\usepackage{graphicx}
\usepackage{graphics}
\usepackage{amssymb}
\usepackage[T1]{fontenc}
\usepackage[utf8]{inputenc}

\title{The exceptional VHE gamma-ray outburst of PKS 1510-089 in May 2016}

\ShortTitle{VHE flare of PKS 1510-089}

\author{\speaker{Michael Zacharias}$^a$, Julian Sitarek$^b$, Dijana Dominis Prester$^c$, Felix Jankowsky$^d$, Elina Lindfors$^e$, Mahmoud Mohamed$^d$, David Sanchez$^f$, Tomislav Terzic$^c$, for the H.E.S.S. and MAGIC Collaborations\\
        $^a$Centre for Space Research, North-West University, 2520 Potchefstroom, South Africa\\
        $^b$University of Lodz, 90236 Lodz, Poland\\
        $^c$Croatian MAGIC Consortium, Rudjer Boskovic Institute, University of Rijeka, University of Split and University of Zagreb, Croatia\\
        $^d$Landessternwarte, Universit\"at Heidelberg, K\"onigstuhl, 69117 Heidelberg, Germany\\
        $^e$Finnish MAGIC Consortium, Tuorla Observatory, University of Turku and Astronomy Division, University of Oulu, Oulu, Finland\\
        $^f$Laboratoire d'Annecy-le-Vieux de Physique des Particules, Universite Savoie Mont-Blanc, CNRS/IN2P3, 74941 Annecy-le-Vieux, France\\
        E-mail: \email{mzacharias.phys@gmail.com}}

\abstract{PKS 1510-089 (z=0.361) is one of only a handful of flat spectrum radio quasars that have been detected at very high energy (VHE, $E>100\,$GeV) gamma rays. It is a very active source across the entire electromagnetic spectrum. VHE observations in May 2016 with H.E.S.S. and MAGIC revealed an exceptionally strong flare, which lasted for less than two nights, and exhibited a peak flux of about 0.8 times the flux of the Crab Nebula above $200\,$GeV. The flare provides the first evidence of intranight variability at VHE in this source. While optical observations with ATOM reveal a counterpart at optical frequencies, {\it Fermi}-LAT observations reveal only low flux variability at high energy (HE, $E>100\,$ MeV) gamma rays. Interestingly, the HE spectral index significantly hardens during the peak of the VHE flare, indicating a strong shift of the peak frequency of the high energy component. Given the expected strong absorption due to the broad-line region, the VHE emission region cannot be located deep within that region.} 

\FullConference{35th International Cosmic Ray Conference – ICRC2017\\
		10-20 July, 2017\\
		Bexco, Busan, Korea}

\begin{document}

\section{Introduction}
Flat spectrum radio quasars (FSRQs) exhibit strong variability across the entire electromagnetic spectrum \cite{tea11,aFea16b}, and show broad optical emission lines (Equivalent Width $>5\,$\AA). FSRQs along with BL Lac objects are the relativistically beamed version of active galactic nuclei, where the jet is closely aligned with the line of sight \cite{br74}. The spectral energy distribution (SED) features two broad humps. According to leptonic models, the low-energy hump is attributed to synchrotron radiation, while the high-energy hump is due to inverse Compton scattering of ambient soft photon fields. In FSRQs the spectral components peak around $1\,$eV and $100\,$MeV, respectively. Hence, FSRQs are notoriously difficult to detect at very-high energy (VHE, $E>100\,$GeV) gamma rays. Currently, six FSRQs have been detected at VHE, and all of them during a flaring event. No quiescent state detection has been achieved so far.



The FSRQ PKS~1510-089 is a particularly active source, and it is monitored in many energy bands. These observations reveal a complex multiwavelength behavior \cite{sea13,sea15,b13,kea16} with changing correlation patterns between different energy bands. This also explains the difficulties in modeling the SED, which has resulted in different interpretations, such as the requirement of at least two radiative emission components to reproduce the inverse Compton component \cite{nea12} or two spatially distinct emission zones \cite{bea14}. While these models are not exclusive, they already point towards the complexity and unpredictable nature of this object.

PKS~1510-089 was detected at VHE with H.E.S.S. during a bright high energy (HE, $E>100\,$MeV) $\gamma$-ray flare in 2009 \cite{Hea13}, and has been observed a few times per year afterwards \cite{zea17}. It has also been observed by MAGIC during another HE high state in 2012 \cite{aMea14}. However, VHE variability could only be established through observations of yet another strong HE flare in 2015 \cite{aMea17,zea16}. In order to study potential connections between VHE variability and flares at other energy bands, observations of PKS~1510-089 at VHE have been intensified in recent years.
These observations resulted in the detection of a strong VHE flare allowing for the first time studies on sub-hour time scales. The results are presented here along with detailed observations at HE $\gamma$ rays and optical energies. 

%
\section{Data analysis and results} \label{sec:ana}
%
%
\subsection{H.E.S.S. and MAGIC analysis} \label{sec:hess}
H.E.S.S. is an array of five Imaging Atmospheric Cherenkov Telescopes (IACTs) located in the Khomas Highland in Namibia at an altitude of about $1800\,$m. The array consists of four 12-m telescopes (CT1-4) and a 28-m telescope (CT5) in the array's center. The optimal energy threshold is $\sim 50\,$GeV. Due to instrumental issues in the observation period (MJD 57535 - 57545), the present analysis only covers events recorded with CT2-4. The reduced number of telescopes, and observation conditions raise the energy threshold of this data set to $\sim 200\,$GeV.

A total number of 31 runs (1 run lasts for about $28\,$min) passed the standard quality selection \cite{aHea06} resulting in a total live time of $13.6\,$hrs. The data set has been analyzed with the Model analysis chain using loose cuts \cite{dnr09}. The results have been cross-checked and verified using the independent reconstruction and analysis chain ImPACT \cite{ph14}.

\begin{figure*}[t]
\begin{minipage}{0.49\linewidth}
\centering 
{\includegraphics[width=0.95\textwidth]{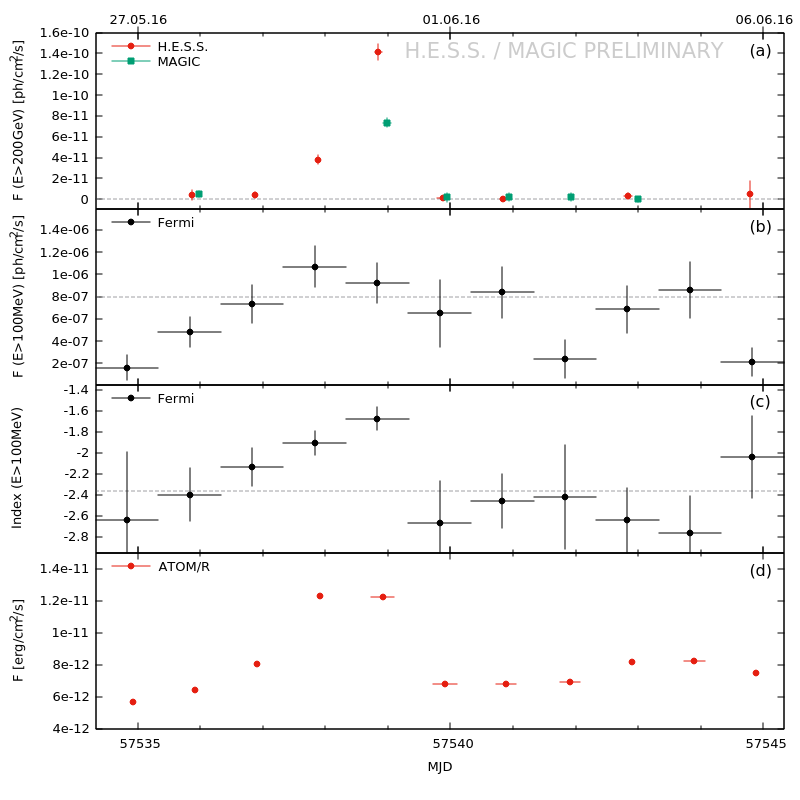}}
\caption{Lightcurves of PKS~1510-089 of the observation period. {\bf (a)} Nightly-averaged VHE lightcurve from H.E.S.S. (red) and MAGIC (green). {\bf (b)} $24\,$hr-average HE lightcurve from {\it Fermi}-LAT. The dashed line marks the $8\,$yr average. {\bf (c)} $24\,$hr-binned HE spectral index assuming a power-law spectrum. The dashed line marks the $8\,$yr average. {\bf (d)} Nightly-averaged optical R-band lightcurve from ATOM.}
\label{fig:mwl_lc_lt}
\end{minipage}
\hspace{\fill}
\begin{minipage}{0.49\linewidth}
\vspace{-0.0cm} \centering 
{\includegraphics[width=0.95\textwidth]{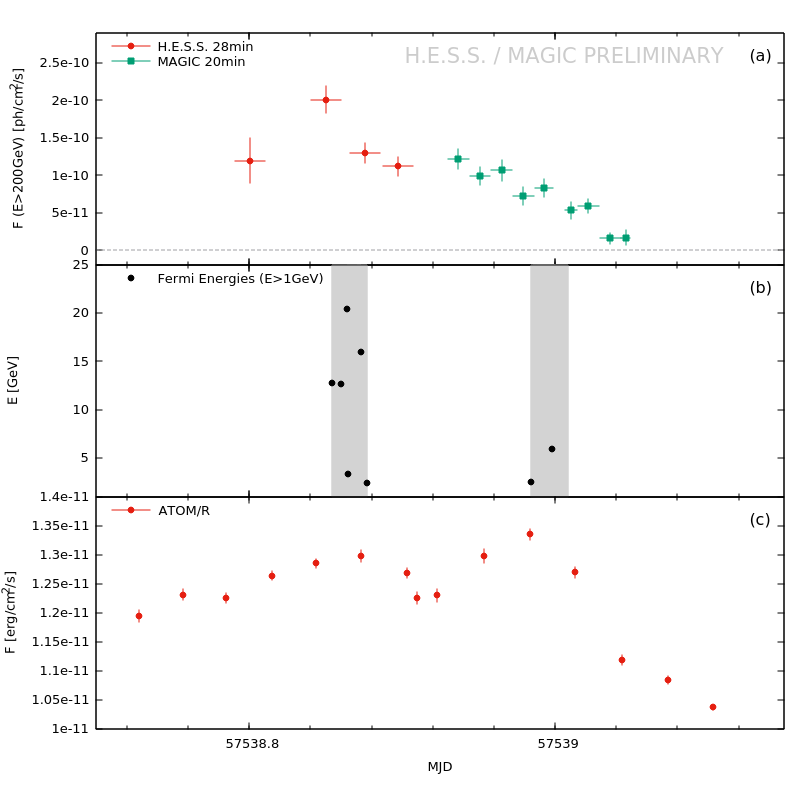}}
\caption{Lightcurves of PKS~1510-089 of the peak night, MJD 57538. {\bf (a)} VHE lightcurves from H.E.S.S. (red) and MAGIC (green) with time-binnings as indicated. {\bf (b)} Detected energies from {\it Fermi} above $1\,$GeV. The gray bands indicate the observable time of PKS~1510-089 for {\it Fermi}. {\bf (c)} Optical R-band lightcurve from ATOM showing individual exposures of about $8\,$min duration.}
\label{fig:mwl_lc_st}
\end{minipage}
\end{figure*}

In the entire observation period the source is detected with a significance of $29.5\sigma$. The average flux during the period above $200\,$GeV is $(2.5 \pm 0.1_{\rm stat} )\times 10^{-11}\,$cm$^{-2}$s$^{-1}$. However, a constant fit to the lightcurve is ruled out on a run-by-run level with more than $10\sigma$. 
The H.E.S.S. lightcurve with nightly averages integrated above $200\,$GeV is shown in red in Fig. \ref{fig:mwl_lc_lt}(a). It is obvious that the source is not significantly detected apart from the two nights beginning on MJD 57537 and 57538. 

On MJD 57538, the source has been observed for 4 runs and a total live time of $1.8\,$hrs. A curved spectrum is preferred with $4.4\sigma$, which is well explained by the absorption from the extragalactic background light (EBL) using the model of \cite{frv08}. The intrinsic spectrum is well described by a power-law 
$ F(E) = N(E_0)\times(E/E_0)^{-\Gamma}$
with normalization $N=(19.0\pm 0.8_{\rm stat})\times 10^{-10}\,$cm$^{-2}$s$^{-1}$TeV$^{-1}$, decorrelation energy $E_0 = 0.268\,$TeV, and photon index $\Gamma = 2.9\pm 0.2_{\rm stat}$. The lightcurve is shown in red in Fig. \ref{fig:mwl_lc_st}(a). The average flux of the night above $200\,$GeV amounts to $(14.3 \pm 0.6_{\rm stat})\times 10^{-11}\,$cm$^{-2}$s$^{-1}$ or $\sim 56\%$ of the Crab nebula flux (or Crab Units, C.U.) above the same energy threshold. However, the flux deviates with $5.4\sigma$ from this average value on a run-wise basis, clearly indicating for the first time VHE intranight variability in PKS~1510-089. The peak flux of the night equals $(20 \pm 1_{\rm stat})\times 10^{-11}\,$cm$^{-2}$s$^{-1}$ or $80\%$ C.U. above $200\,$GeV. There is no evidence for intra-run variability.


%
%

%
%
MAGIC is a system of two IACTs with a mirror dish diameter of $17\,$m each, located in Canary Island of La Palma, at the height of 2200 m a.s.l. \cite{al16a}. As PKS~1510-089 is a southern source, only observable by MAGIC at zenith angle above $38^\circ$, the corresponding trigger threshold, defined as a peak of the MC energy distribution for a Crab-like spectrum, is $\gtrsim90\,$GeV \cite{al16b}. Note that such a peak is rather broad, and especially for softer than Crab sources it is possible to reconstruct source flux below this value.

The last observations of PKS~1510-089 before the flare took place on MJD 57535. During the night of the flare MAGIC observed PKS~1510-089 for $2.53\,$h, and nightly follow up observations were performed until MJD 57542, resulting in a total live time of $7.5\,$h. The data has been reduced and analyzed using MARS, the standard analysis package of MAGIC \cite{za13, al16b}. 

During the investigated period, only in the night beginning on MJD 57538 a significant signal has been observed. The average flux above $200\,$GeV on this night equals $(7.36\pm0.40_{\rm stat})\times 10^{-11}\, \mathrm{cm^{-2}s^{-1}}$ (corresponding to $\sim 32\%$ C.U. above the same energy threshold). The flux increased by a factor of $\gtrsim 5$ with respect to the flux upper limits measured on neighboring nights, as can be seen by the green points in Fig. \ref{fig:mwl_lc_lt}(a). 
The intranight light curve ($20\,$min bins), as shown by the green points in Fig. \ref{fig:mwl_lc_st}(a), exhibits clear variability. A constant fit can be excluded with more than $10\sigma$. 
The flux drops from $\sim50\%$ C.U. at the beginning of the observation down to $\sim 7.5\%$ C.U. at the end of the night. 

The observed spectrum of the night of the flare was reconstructed between $60$ and $700\,$GeV. Its curvature is well explained by the EBL absorption. Using the EBL model by \cite{frv08}, the intrinsic spectrum can be described by a power-law with normalization $N=(34.7\pm 1.5_{\rm stat})\times 10^{-10}\,$cm$^{-2}$s$^{-1}$TeV$^{-1}$, decorrelation energy $E_0 = 0.175\,$TeV and an index of $\Gamma = 3.37\pm 0.09_{\rm stat}$.

Only statistical errors have been quoted. For both observatories, the systematic uncertainty on the energy scale is $15\%$.

%
\subsection{Multiwavelength analysis} \label{sec:fermi}
HE $\gamma$-ray data has been taken from {\it Fermi}-LAT observations, which covers the full sky every $3\,$hrs in the energy range above $20\,$MeV \cite{aFea09}. An analysis of the (publicly available) Pass 8 SOURCE class events above an energy of $100\,$MeV was performed for a Region of Interest (RoI) of $15^{\circ}$ radius centered at the position of PKS~1510–089. All sources within the RoI (and $7^{\circ}$ beyond) pertaining to the 3FGL catalog have been accounted for in the likelihood analysis. In order to reduce contamination from the Earth Limb, a zenith angle cut of $<90^{\circ}$ was applied. The analysis was performed with the ScienceTools software package version v10r0p5 using the \textsc{P8R2\_SOURCE\_V6} 
instrument response function and the \textsc{GLL\_IEM\_v06} and \textsc{ISO\_P8R2\_SOURCE\_V6\_v06} models 
for the Galactic and isotropic diffuse emission \cite{aFea16a}, respectively.

In order to produce a finely binned ($24\,$hr bins) lightcurve, a power-law model for the spectrum was assumed with the parameters left free to vary from bin to bin. The flux points have been centered around the VHE observation windows. The resulting lightcurve is shown in Fig. \ref{fig:mwl_lc_lt}(b). Given that the average flux of PKS~1510-089 in this energy band (after about $8\,$years of {\it Fermi} observations) is $8.8\times 10^{-7}\,$ph$\,$cm$^{-2}$s$^{-1}$, the flux variation is not particularly strong, and by no means matches HE outbursts already detected in this source \cite{sea13,aMea17}. On the other hand, the photon index shows significant hardening of the HE spectrum during the VHE flare (see Fig. \ref{fig:mwl_lc_lt}(c)), reaching an index as low as $1.7\pm 0.1$, and even harder on shorter time scales (see below). The average power-law index of PKS~1510-089 in the HE band since 2008 is $2.368\pm0.004$. This indicates that the flare mainly influenced the highest energies, while lower energies were not particularly affected. Hence, the peak frequency of the high energy component was significantly shifted to higher energies during the outburst.

The hardening of the spectrum has been analyzed further by considering the arrival times of the most energetic photons ($E>1\,$GeV) during the peak of the VHE flare. As can be seen in Fig. \ref{fig:mwl_lc_st}(b), the most energetic photons with energies between $10\,$GeV and $25\,$GeV arrived during the H.E.S.S. observation window, i.e. near the peak of the VHE lightcurve, while during the MAGIC observation window the photons registered by {\it Fermi}-LAT had energies less than $10\,$GeV. The resulting spectra are shown in Fig. \ref{fig:spec_gamma}. The butterflies have been computed for the precise H.E.S.S. and MAGIC observation times (MJD 57538.79 - 57538.91 and MJD 57538.93 - 57539.05, respectively) and have been cut at the highest detected photon energy in the respective windows. The two HE spectra are marginally consistent. The photon index during the H.E.S.S. window equals $\Gamma = 1.4\pm0.2$, while it is $\Gamma = 1.7\pm0.2$ during the MAGIC window. 

%
%
Optical R-band data has been collected with ATOM, a 75 cm optical telescope located on the H.E.S.S. site \cite{hea04}. Operating since 2005, ATOM provides optical monitoring of $\gamma$-ray sources and has observed PKS~1510-089 with high cadence. The magnitudes of each flux point have been derived with differential photometry using five comparison stars in the same field of view. The resulting fluxes have been corrected for Galactic extinction.

After the beginning of the VHE flare the frequency of observations was increased further to several exposures per night resulting in a detailed lightcurve. Fig. \ref{fig:mwl_lc_lt}(d) shows the nightly averaged optical lightcurve, while Fig. \ref{fig:mwl_lc_st}(c) displays individual exposures for the flare night, MJD 57538. It is apparent that the optical flux exhibited an outburst, even though the flux at the time of the flare was several times below the historical high state in July 2015 \cite{jea15}. 

%
\section{The $\gamma$-ray flare} \label{sec:corstd}
%
%
%
%
\begin{figure*}[t]
\begin{minipage}{0.62\linewidth}
\centering 
\includegraphics[width=0.80\textwidth]{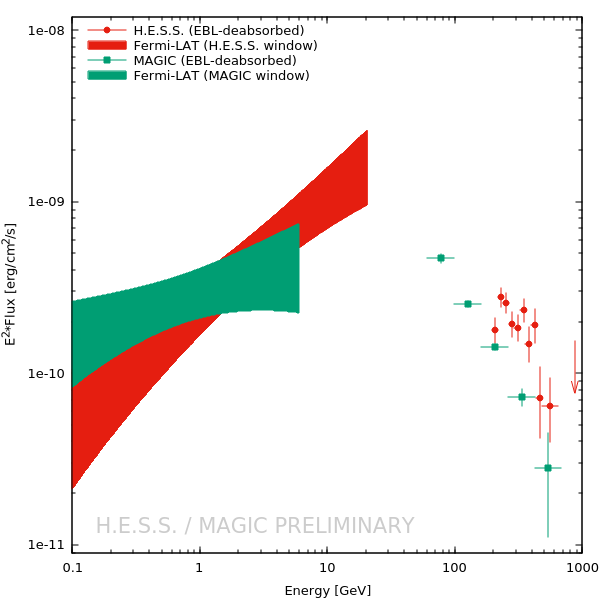}
\end{minipage}
\hspace{\fill}
\begin{minipage}{0.37\linewidth}
\caption{Observed HE and VHE $\gamma$-ray spectra of PKS~1510-089 during the peak night. The {\it Fermi} butterflies are integrated over the precise H.E.S.S. (red) and MAGIC (green) observation windows and are plotted until the highest detected photon energy. The VHE spectra are EBL de-absorbed, with the H.E.S.S. spectrum in red and the MAGIC spectrum in green.}
\label{fig:spec_gamma}
\end{minipage}
\end{figure*} 
%
In Fig. \ref{fig:spec_gamma} the HE and VHE spectra of the peak night (MJD 57538) are plotted. The apparent softening of the HE spectrum from one observation window to the other, despite being not statistically significant, is probably influenced by the (non-)detection of photons with energies exceeding $10\,$GeV during the observation windows (see Fig. \ref{fig:mwl_lc_st}(b)). The peak of the $\gamma$-ray spectrum is located somewhere between $10$ and $60\,$GeV. Compared to the usual location below $1\,$GeV \cite{bea14}, the peak position shifted by more than a factor of $10$ to higher energies.  

The EBL-deabsorbed VHE spectra are devoid of any curvature. Interestingly, the break between the HE and VHE spectra does not change significantly between the observation windows. During the H.E.S.S. window it is $\Delta\Gamma = 1.5\pm0.3_{\rm stat}$, while during the MAGIC window it equals $\Delta\Gamma = 1.7\pm0.2_{\rm stat}$. Hence, whatever is causing the break, it is steady during the evolution of the flare. While this might be a Klein-Nishina break, one could interpret this effect as being due to absorption by a soft photon field located close to the central engine, like the broad-line region (BLR). One can conservatively estimate an upper limit on the degree of absorption by assuming that the spectrum detected by {\it Fermi} intrinsically continues unchanged into the VHE domain \cite{zea16}. The degree of absorption $\tau$ can then be derived by
\begin{eqnarray}
 \tau = \ln{\frac{F_{extra}}{F_{obs}}} \label{eq:tau},
\end{eqnarray}
where $F_{extra}$ is the extrapolated flux, and $F_{obs}$ is the observed flux.

Without going into details \cite{be16,zea16}, the calculation for the H.E.S.S. data set give a maximum value of $\tau$ for the highest energies of $\tau=5.4\pm0.9_{\rm stat}$, while the MAGIC data gives a maximum value of $\tau=3.9\pm1.4_{\rm stat}$. These estimates agree within errors. Assuming the absorption is due to the BLR,\footnote{While the maximum absorption due to the $H\alpha$ line in the BLR should occur at $\sim 25\,$GeV, the absorption in the VHE domain could be caused by a thermal spectrum with a temperature of a few thousand Kelvin.} the absorption values can be translated into a minimum distance of the emission region from the black hole. The emission region could be located at roughly $3\times 10^{17}\,$cm from the black hole within the inner boundary of the BLR, if the latter is assumed to be a spherical shell. It could be located even further in, if the pairs created from the absorption process launch an electron-positron cascade \cite{sb10,rb10}.

%
%
An independent estimate of the source location can be derived using the minimum variability time scale between subsequent flux measurements, which is defined as \cite{zea99}
\begin{eqnarray}
 t_{var} = \frac{F_i+F_{i+1}}{2} \frac{t_{i+1}-t_{i}}{|F_{i+1}-F_{i}|} \label{eq:tvar}.
\end{eqnarray}
The fastest variability occurs between the seventh and eighths point of the MAGIC window with $t_{var}=18\pm5_{\rm stat}\,$min. During the H.E.S.S. observation window, the minimum variability time scale is $t_{var}=83\pm26_{\rm stat}\,$min between the second and third point.

The variability constrains the size of the emission region to $R\sim 10^{15}(\delta/50)\,$cm, where $\delta$ is the Doppler factor of the emission region. If the emission region fills the entire width of the conical jet with opening angle $\theta_j\sim 0.2/\Gamma_b$ \cite{cea13}, the distance from the black hole roughly becomes $d\sim R/\theta_j \sim 2.5\times 10^{17}\,$cm, which agrees with the estimate from the absorption. Here we assumed that the bulk Lorentz factor $\Gamma_b$ equals the Doppler factor.
However, if the break in the $\gamma$-ray spectrum is not primarily caused by absorption, the emission region is located further away from the black hole, and could no longer fill the entire widths of the jet.

%
\section{The optical flare}
The R-band observations with ATOM covered the entire night of the flare, Fig. \ref{fig:mwl_lc_st}(c). While the peak of the optical flux is more than a factor 3 above the quiescence level of PKS~1510-089, the variability during the night is only on the order of $30\%$. This is in strong contrast to the VHE variability, which is more than a factor 10. Furthermore, the optical lightcurve exhibits a double-peaked structure, which is also not seen in the VHE lightcurve. Interestingly, the significant drop shortly after midnight as seen in VHE lightcurve of MAGIC, also seems to take place in the optical band around the same time. Apart from that, the evolution of the optical flare differs significantly from the VHE flare.

This and the lack of additional data inhibit strong conclusions about the cause of the flare without in-depth modeling. Nevertheless, whatever caused the huge outburst in the VHE, can have left only a small mark on the optical synchrotron component.

%
\section{Summary}
The May-2016-flare in PKS~1510-089 was a very unusual event, since the tremendous outburst in the VHE $\gamma$ rays has only mild counterparts in the HE $\gamma$ rays and the optical R-band. The flux in the VHE band varied by more than a factor 10 in a single night, while in the same night the optical flux varied by only $30\%$. Surprisingly, the integrated flux in the HE band did not vary at all, even though the HE spectrum hardened significantly, with the photon index reaching values as low as $1.7\pm0.1$ in $24\,$h bins. Hence, the peak of the inverse Compton component shifted by more than a factor $10$ to higher energies.

The VHE flux varied on time scales of less than an hour, proving for the first time intra-night variability in the VHE domain in this source. The spectra measured by H.E.S.S. and MAGIC were significantly curved, which can be fully explained by EBL absorption. The intrinsic VHE spectra are compatible with power-laws. {\it Fermi}-LAT spectra have been derived for the precise H.E.S.S. and MAGIC observation windows, which reveal that during the peak of the VHE lightcurve (during the H.E.S.S. observations) {\it Fermi}-LAT measured a harder power-law and detected higher energetic photons than during the MAGIC observation window, where the VHE flare was declining. Interestingly, the break between the HE and VHE spectrum remained constant within errors over the duration of the flare. 
The break in the $\gamma$-ray spectrum could result from absorption by soft BLR photons. Along with estimates from the fast variability, this places the emission region at the inner edge of the BLR, even though larger distances cannot be excluded.

Contrary to the single-peaked lightcurve in the VHE domain, the optical lightcurve from ATOM exhibited a double-peaked structure. Apparently, whatever caused the sharp VHE flare, must have had only a mild, yet somewhat different impact on the optical synchrotron flux.

The unusual event in PKS~1510-089 further supports efforts to conduct blazar monitoring in as many bands as possible, since blazars never stop to surprise and challenge our theoretical description of them.

\section*{Acknowledgements}
\noindent H.E.S.S. gratefully acknowledges financial support from the agencies and organizations listed at
\href{https://www.mpi-hd.mpg.de/hfm/HESS/pages/publications/auxiliary/HESS-Acknowledgements-ICRC2017.html}{\footnotesize{https://www.mpi-hd.mpg.de/hfm/HESS/pages/publications/auxiliary/HESS-Acknowledgements-ICRC2017.html}}

\noindent MAGIC gratefully acknowledges financial support from the agencies and organizations listed at
\href{https://magic.mpp.mpg.de/acknowledgements_19_05_2017.html}{\footnotesize{https://magic.mpp.mpg.de/acknowledgements\_19\_05\_2017.html}}.

\end{document}